\documentclass[twocolumn,showpacs,preprintnumbers,amsmath,amssymb]{revtex4}

\usepackage{psfig} 
\usepackage{stmaryrd} 
\usepackage{graphicx} 

\begin{document}

\preprint{APS/123-QED}

\title{Product Distribution Field Theory}

\author{David H. Wolpert}
\affiliation{NASA Ames Research Center, Moffett Field, CA, 94035, USA\\
{\tt dhw@email.arc.nasa.gov}}

\begin{abstract} This paper presents a novel way to approximate a
distribution governing a system of coupled particles with a product of
independent distributions.  The approach is an extension of mean field
theory that allows the independent distributions to live in a
different space from the system, and thereby capture statistical
dependencies in that system. It also allows different Hamiltonians for
each independent distribution, to facilitate Monte Carlo estimation of
those distributions. The approach leads to a novel energy-minimization
algorithm in which each coordinate Monte Carlo estimates an associated
spectrum, and then independently sets its state by sampling a
Boltzmann distribution across that spectrum.  It can also be used for
high-dimensional numerical integration, (constrained) combinatorial
optimization, and adaptive distributed control.  This approach also
provides a simple, physics-based derivation of the powerful
approximate energy-minimization algorithms semi-formally derived
in~\cite{wowh00, wotu02c, wolp03a}. In addition it suggests many
improvements to those algorithms, and motivates a new (bounded
rationality) game theory equilibrium concept.  \end{abstract}



\maketitle              

\nocite{chjo02, kono03_short}

\section{Introduction} \label{sec:intro} Mean Field Approximations
(MFA's) have many applications outside of physics, combinatorial
optimization by mean-field annealing being perhaps the most prominent
of them~\cite{pean87,nish01}. One problem with MFA's is that because
they treat all particles as independent, they can be a poor
approximation of the desired distribution. Another is that especially
in large systems with complex Hamiltonians, they can be expensive to
evaluate.

We extend MFA's by introducing coordinate transformations that allow
us to convert a coupled distribution into a decoupled one. This allows
us to improve the accuracy of the MFA. It also allows us to employ
Monte Carlo mean-field annealing for optimal adaptive control. We also
introduce separate Hamiltonians for each particle. This allows us to
exploit recent results in collectives theory~\cite{wolp03a} to speed
up MFA evaluations. This extended MFA justifies heuristics recently
found to beat simulated annealing in adaptive control problems by
orders of magnitude~\cite{wowh00, wotu02c, wotu03a}. It also provides
novel ways to do (constrained) combinatorial optimization, a novel
(bounded rational) game theory equilibrium
concept~\cite{auha92,pisl02}, and a way to calculate such equilibria.

\section{Self-consistent Distributions}

Let $\zeta$ be a space with elements $z$ and cardinality $|\zeta|$.  A
{\it semi-coordinate system} of $\zeta$, $\Xi$, is a space $\xi$
having elements $x = (x_1 \in \xi_1, \ldots, x_{M} \in \xi_{M})$
together with a surjective map from $\xi$ to $\zeta$, written as
$\zeta(.)$.  $\zeta(.)$ need not be invertible; the (semi-)coordinates
of a point $x \in \xi$ may not be unique. No a priori restriction is
made on whether $\xi$ and $\zeta$ are countable, uncountable,
time-extended, etc.

The space of possible probability distributions (or density functions,
as the case may be) over $\xi$ is $\cal{P}^\dagger$. Any $p \in
{\cal{P}}^\dagger$ induces a distribution over $\zeta$, $p(z) = \int
dx \; p(x) \delta(z - \zeta(x))$. (We implicitly assume integral
measures and delta functions that match $\xi$.)  Expressions like
``$p(\xi) = p'(\xi)$'', ``$p_i = p'_i$" and ``$p_{(i)} = p'_{(i)}$'',
where $p, p' \in {\cal{P}}^\dagger$, mean $\forall x \in \xi$, $p(x) =
p'(x)$, $p(x_i) = p'(x_i)$, and $p(x_j) = p'(x_j) \; \forall j \ne i$,
respectively.  $\cal{P}$ is the set of all product distributions over
$\xi$, i.e., the submanifold of all $p \in {\cal{P}}^\dagger$ obeying
$p(\xi) = \prod_i p(\xi_i)$. We will use the obvious parameterization
of elements of $\cal{P}$ as the vectors of their marginal
distributions, written $q = (q_1(\xi_1), \ldots, q_M(\xi_M))$. Note
that changing $\zeta(.)$ with $\xi$ fixed will change the manifold
$\cal{P}$ in general but won't affect $\cal{Q}$, the space of all $q$.

We are interested in {\it self-consistent} $q$ that obey $q_i(\xi_i) =
A_i(\xi_i, q_{_{(i)}}) \; \forall i$ for some functions $\{A_i\}$. By
Brouwer's theorem, for any smooth \{$A_i$\}, the map $q \rightarrow
\{A_i(\xi_i, q_{_{(i)}})\}$ (which we call {\it parallel Brouwer
updating}) has at least one fixed point. Here we consider such $A_i$
that set $q_i$ by minimizing a functional of $(q_i, q_{_{(i)}})$. As
an example, that functional might be the same for all $i$, for each
one measuring the error of $(q_i, q_{_{(i)}})$ as an approximation of
some $p \in {\cal{P}}^\dagger$. In this case the fixed point(s) of the
\{$A_i$\} are locally optimal approximations in $\cal{P}$ of that $p$.

Write the cross entropy from $p$ to $p'$ as $S(p \; || \; p') \equiv
-\int dx \; p(x) {\mbox{ln}}[\frac{p'(x)}{\mu(x)}]$, where as usual
$\mu$ is a prior probability over $\xi$ that ensures the argument of
the logarithm is unitless. So the entropy of $p$ is just $S(p \; || \;
p) \equiv S(p)$, and the Kullback-Leibler distance from $p$ to $p'$ is
$KL(p \; || \; p') \equiv S(p \; || \; p') - S(p)$~\cite{coth91}.
Next define {\it private Hamiltonians} $h_{i,j}$ where $\zeta(x) =
\zeta(x') \Rightarrow h_{i,j}(x) = h_{i,j}(x') \; \forall i, j, x$ and
$x'$, so expected values of a Hamiltonian have the same value in
$\zeta$ and $\xi$.  We write $h_i(\xi) \equiv \sum_{j=1}^{m_i}
\beta_{i,j} h_{i,j}(\xi)$ where the temperatures $\{\beta_{i,j}\}$ are
usually non-negative and write the associated free energies as
${F}_{h_i}(p) \equiv \int dx \; p(x) h_{i}(x) \; - \; S(p)$.  So
argmin$_{q'_i \in {\cal{Q}}_i} {F}_{h_i}(q'_i, q_{_{(i)}})$ is the
$q'_i$ maximizing the entropy of $q'_i(\xi_i) \prod_{k \ne i}
q_{k}(\xi_{k})$, subject to expected values of $i$'s private
Hamiltonians implicit in $i$'s temperatures. The associated fixed
points obey

$ $

$
q_i^{g}(\xi_i) \; = \;
\frac{\mu_i(\xi_i)}{ N_i({[h_i]}_{i,q^g})} 
e^{-{[h_i]}_{i,q^g}(\xi_{i})}
$

$ $

\noindent where $N_i([U]_{i,p})$ is the partition function and the
bracket notation indicates expectation conditioned on the value of the
subscripted coordinate: for any $U$ and $p \in {\cal{P}}^\dagger$,
$[U]_{i,p}(\xi_{i}) \equiv \int d{x_{(i)}} U(\xi_i, x_{(i)}) p(x_{(i)}
\mid \xi_i)$.

We can choose \{$A_i$\} based on objective functions other than the
free energy; the important thing is to incorporate a concave function
like entropy, to avoid the expensive process of checking the borders
of $\cal{Q}$ for fixed points. For example, all the local minima of
${F}_{h_i}(q)$ over ${\cal{Q}}$ are interior to ${\cal{Q}}$.
Moreover, for any fixed $q_{_{(i)}}$, there is only one local minimum
over ${\cal{Q}}_i$.

The following six examples all have uniform $\mu$.  $\xi = \zeta$ (as
in conventional MFA's) in all but the fifth example:

$ $

{\it Example 1}: Let $m_i = 1 \; \forall i$. Then at $q^g$
each coordinate $i$ obeys its own canonical ensemble and is
independent of the values of the other coordinates. It is coupled to
its own heat bath, with its own effective Hamiltonian set by the
distributions over the other coordinates, ${[h_i]}_{i,q^g}(\xi_i)$.
In contrast, if $\forall i, m_i = 2$ and $h_{i,2} = (h_{i,1})^2$, each
$q^{g}_i(\xi_i)$ is a Gaussian over the values ${[h_i]}_{i,1}(\xi_i)$
rather than a Boltzmann distribution over them.  In this case
$\beta_{i,1}$ and $\beta_{i,2}$ specify the mean and variance of
$h_{i,1}$.

{\it Example 2}: We have a {\it team game} when $\forall i,
i'$, and $j$, $m_i = m_{i'}$, and $h_{i} = h_{i'}$.  In a team game we
have a single shared objective function trading off entropy and
expected values of the Hamiltonians.  For $m = 1$ we define the {\it
world Hamiltonian} $H$ as the single shared private Hamiltonian. For
this the minimizer of the free energy over all ${\cal{P}}^\dagger$ is
the joint canonical ensemble distribution, ${\bf{p}}^{\beta
H}$~\cite{jayn57}. In contrast, $q^{\beta H}$, the minimizer over
${\cal{P}}$, is an MFA for ${\bf{p}}^{\beta H}$.  Since $\zeta$ is
arbitrary, by gradually increasing $\beta$ one can converge on a $q$
that is a delta function about the $z$ that minimizes $H(z)$. Since
$\zeta$ is arbitrary, this optimization algorithm can be applied for
any underlying space one wishes to search. In particular, parallel
Brouwer updating with gradually increasing $\beta$ is equivalent to
mean-field annealing~\cite{pean87}.

\noindent In general $q^{\beta H}(\xi_i) \ne {\bf{p}}^{\beta
H}(\xi_i)$, the marginal of ${\bf{p}}^{\beta H}(\xi_i)$. However the
$q^{\beta{H}}$ with the lowest free energy is the $q$ that best
approximates ${\bf{p}}^{\beta{H}}$, as measured by $KL(q^{\beta H},
{\bf{p}}^{\beta H})$ To have the fixed points equal the marginals we
must instead have the $\{A_i\}$ minimize $KL({\bf{p}}^{\beta H},
q^{\beta H})$.)

{\it Example 3:} One way to enforce a constraint $f(z) = 0$ is via a
generalization of penalty functions.  Say we have an $m=1$ team game
with world Hamiltonian $H$.  Choose the objective function $\int dx
p(x) [\beta H(x) + \beta_f [f(\zeta(x))]^2] - S(p)$. Then for $\beta_f
\rightarrow \infty$ our constraint will be enforced, if possible.
Often that solution is on ${\cal{Q}}$'s border though, which slows the
search --- finite $\beta_f$ moves the solution(s) to the interior of
${\cal{Q}}$, by weakening the constraint. Alternatively, choose the
objective function $\int dx p(x) (\beta H(x) + \lambda
[f(\zeta(x))]^2) + \gamma \lambda^2 - S(p)$, where $\gamma \ge 0 $ is
a constant.  Minimizing over both $p \in {\cal{Q}}$ and $\lambda \in
\mathbb{R}$ forces $f(z) = 0$ if $\gamma = 0$, while nonzero
$\gamma_j$ weakens the constraint but forces the fixed point $q$(s)
inside ${\cal{Q}}$.

{\it Example 4}: Consider a team game with $m = m'+1$ and
$h_{j}(x) = h_{j}(x_j) \in {\mathbb{N}}^+ \; \forall j \le m'$. Here
$p^{g}$ is an MFA for the grand canonical ensemble of a system with
$m'$ particle types and Hamiltonian $h_{m'+1}$: $x_i$ encodes the
states of all type $i$ particles, and $h_{i}$ counts their
number~\cite{jayn57}.

{\it Example 5}: Say $\zeta(.)$ is bijective but $\xi \ne
\zeta$. Then $\zeta(.)$ is akin to a rotation, in that each $\xi_i$
couples multiple components of $\zeta$. More generally, $|\xi| \ne
|\zeta|$ means $\xi$ is an $M$-dimensional representation of an $M' <
M$-dimension system.  This allows arbitrary distributions over $\zeta$
to be expressed as a product distribution.

\noindent As an illustration, say $\zeta$ is two-dimensional with
elements $(z_1, z_2)$. Take $\xi_1 = \zeta_1$, and have an additional
$\xi$ coordinate for each separate value of $z_1$, which we write as
$\xi_{z_1}$. So there are $1+|\zeta_{1}|$ coordinates altogether. Have
each $\xi_{z_{1}}$ contain $|\zeta_2|$ elements, i.e., $|\zeta_2|$
separate subsets of $\zeta$, and label those subsets by the values
$z_2 \in \zeta_2$.  Have $\zeta(x_1, \ldots, x_{1 + |\zeta_1|}) =
(x_1, x_{x_1})$, i.e., $z_1 = x_1$ and $z_2 = x_{z_1}$. So we can
write $p(z_1) = \sum_{z_2} p(z_1, z_2) = \sum_{z_2} \sum_{x: x_1 =
z_1, x_{z_1} = z_2} q(x) = q_{x_1}(z_1)$. Therefore $p(z_2 \mid z_{1})
= q(\xi_{z_{1}} = z_2 \mid \xi_1 = z_1)$. So $p(z_2 \mid z_{1}) =
q_{z_1}(z_2)$.

\noindent This allows us to do optimal distributed adaptive control in
$\zeta$ with an MFA over $\xi$. Let $\zeta_2$ be the state of the
plant one wants to control, $\zeta_{1}$ the control variable you can
set, and ${\cal{H}}(\zeta_2)$ the objective function one wants to
minimize. The goal in optimal control is to find argmin$_{z_{1}}
E({\cal{H}} \mid z_{1})$. Have the $q_{i \ge 2}$ fixed to $P(\zeta_2
\mid \zeta_{1})$, the distribution relating the plant variable and the
control variables. Take $m_1 = 1$, $h_{1,1}(\xi) =
{\cal{H}}(\zeta(\xi))$, and redefine $F_{h_1}$ to involve the entropy
of just $\xi_1$ rather than all of $\xi$.  Anneal $\beta \rightarrow
\infty$, so that $q_1(\xi_1)$ becomes a single delta function. Write
the resultant $q^{\beta H}(\xi)$ as $ \delta_{\xi_1, x_1} \prod_{i \ge
2} q_i(\xi_i)$. This distribution minimizes $\sum_{x'} h_{1,1}(x')
\delta_{x'_1, x_1} \prod_{i \ge 2} q_i(x'_i) = \sum_{x'} h_{1,1}(x')
q_{x_1}(x'_{x_1}) = \sum_{z} {\cal{H}}(z) p(z_2 \mid z_1 = x_1)$. So
$z_1 = x_1$ solves our optimal control problem.

{\it Example 6}: Formally, when $m_i = 1 \; \forall i$ and $\xi =
\zeta$, each coordinate is a player in a non-cooperative game
~\cite{auha92,pisl02}, with $x$ the players' joint move and
\{$-h_{i,1}$\} their payoff functions. As the $\beta_i \rightarrow
\infty$, $q^{\{\beta_i h_{i,1}\}}$ becomes a (normal form) mixed
strategy Nash equilibria.

\noindent Now consider the general bounded rational (i.e., non-Nash equilibrium)
scenario. Say the players are told the entropy of the joint system,
and each finds its best possible mixed strategy subject to the other
players' strategies and entropy value. Then the system is at a $q^g$.

\noindent As an alternative interpretation of $q^g$, say we have a {\it
rationality} functional $R(U, q_i)$ that measures how peaked any $q_i$
is about argmax$_{x_i} U(x_i)$ for any $U : \xi_i \rightarrow
\mathbb{R}$. We require that $R(U, q_i) = \beta$ if $q_i(\xi_i)
\propto \mu(\xi_i) e^{-\beta U(\xi_i)}$, and that the $q_i$ satisfying
$R(U, q_i) = \beta$ that has maximal entropy is $\mu(\xi_i) e^{-\beta
U(\xi_i)} / N_i(U)$.  Now say we are told the \{$\beta^*_i$\}\}, the
rationalities of the $M$ players for their associated effective
Hamiltonians \{${[h_{i,1}]}_{i,q}$\}. Then the information-theoretic
optimal estimate for the associated $q$ is the minimizer over $q$ and
the \{$\lambda_i$\} of the free energy $\sum_i \lambda_i
[f_i(R({[h_{i,1}]}_{i,q}, q_i)) - f_i(\beta^*_i)] - S(q)$ for any
monotonically increasing functions $\{f_i\}$~\cite{jayn57}. At any
local minimum of this free energy $q = q^g$ with $\beta_i = \beta^*_i
\; \forall i$.

\noindent This use of rationality functionals expresses the
bounded-rational solution to any non-team game as the minimizer of a
single objective function whose local minima are all interior to
$\cal{Q}$.  As an illustration, choose $f_i(\beta)$ to be the ideal
expected Hamiltonian $\partial_{\beta} {\mbox{ln}}(N_i(\beta
{[h_{i,1}]}_{i,q}))$, and $R(U, q_i) \equiv {\mbox{argmin}}_\beta \;
KL(q_i \; || \; \frac{e^{-\beta U}}{N_i(\beta U)})$. For such a choice
$f_i(R({[h_{i,1}]}_{i,q}, q_i))$ is the actual expected Hamiltonian,
$\int dx_i \; q_i(x_i) {[h_{i,1}]}_{i,q}(x_i)$.

\section{Finding Fixed Points}

When we have an $m=1$ team game we can use variants of gradient
descent to find minima of the (single) free energy. Another approach
for such scenarios is parallel Brouwer updating. More generally,
define the {\it free energy gap} at $q$ for coordinate $i$ as
ln$[N_i({[h_i]}_{i,q^g})] + \int dx_i q_i(x_i) {[h_i]}_{i,q^g}(x_i) +
\int dx_i q_i(x_i) {\mbox{ln}}\frac{q_i(x_i)}{\mu_i(x_i)}$.  This is
how much ${F}_{h_i}$ is reduced if only $q_i$ undergoes the Brouwer
update.  Define {\it serial} Brouwer updating as only updating one
$q_i$ at a time. In an $m=1$ team game, any such update must reduce
${F}_{\beta{H}}(p)$, in contrast to the case with parallel Brouwer
updating.  In {\it greedy} serial Brouwer updating, instead of cycling
through all $i$, at each iteration we update only the coordinate with
the largest gap; this maximizes the free energy drop in that update.

A practical difficulty with these schemes for finding fixed points is
that evaluating ${[h_i]}_{i,q^g}$ can be very difficult in large
systems.  An alternative is to use Monte Carlo simple sampling to get
estimates of the effective Hamiltonians, and use those to update $p$.
In this scheme, given a $q$ at iteration $t$, each $x_i$ is separately
set by randomly sampling $q_i(t)$, thereby generating $x(t)$. Next the
pair $(x_i(t), h_i(x(t)))$ is combined with previous pairs and the
update rule to set $q_i(t+1)$, and then the process repeats.

To simplify the analysis, consider simple gradient descent of the free
energy for the $m=1$ team game. Have $p$ be constant through each
successive block of $\tau$ timesteps, updating only when we go from
some block $m$ to block $m + 1$, with the update based on observations
during block $m$.  Say we have a team game and are at a
block-transition, $t = \tau m + 1$, and let $n_i(t) \in \nu_i(t)$ be
all information the algorithm controlling $q_i$ has at that time,
including the associated Monte Carlo samples. So we have a posterior
conditional distribution of possible gradient descent directions
$P(\vec{F}^i_{{H},\beta}(q(t)) \mid n_i(t))$, where $\vec{F}^i_{H,
\beta}(q(t))$ is the components involving coordinate $i$ of the
projection onto ${\cal{P}}$ of the free energy gradient $\nabla
F_{{H}, \beta}(q(t))$.

In gradient descent updating, that distribution should set the vector
to add to $q_i(t)$ to get $q_i(t+1)$.  More precisely, since agent $i$
knows $q_i(t)$, presuming quadratic loss reflects quality of the
update, the Bayes-optimal estimate of the gradient is the posterior
expected gradient, $\int d[q_{_{(i)}}(t)] \; P(q_{_{(i)}}(t) \mid n_i)
\times \vec{F}^i_{H, \beta}(q(t))$. Expanding, the $x_i$ component of
$\vec{F}^i_{H, \beta}(q)$ is $u_i(x_i) - \sum_{x_i} u_i(x_i) /
|{\xi}_i|$, where $u_i(x_i) \equiv \beta {[H]}_{i,q^g}(x_i) \;+$
ln$[q(x_i)]$.  Rather than evaluate the integral of
${[H]}_{i,q^g}(\xi_i)$, we can use a maximum likelihood estimator,
i.e., replace that integral with $({\hat{H}})_{i,{n_i}}(\xi_i)$, the
average of the observed $H$ values over the $\tau_{x_i}$ instances (of
the just-completed block) when $\xi_i = x_i$.

Unfortunately, often in very large systems the convergence of
$(\hat{H})_{i,n_i}(\xi_i)$ is very slow, since the distribution sampled by
the Monte Carlo to produce $n_i$ is very broad.  To address this,
posit that the differences \{$(\hat{H})_{i,q^g}(x_i) -
(\hat{H})_{i,q^g}(x'_i), \; x_i, x'_i \in \xi_i$\} are unchanged when
one replaces $H$ with some $h_i$.  This means that $q_i$(t)
is unchanged by that replacement. The set of all $h_i$ guaranteed to
have this character, regardless of the form of $q(x_{i},
\xi_{_{(i)}})$, is the set of all {\it difference Hamiltonians},
$h_i({\xi}) = {H}({\xi}) - D_i({\xi}_{_{(i)}})$ for some function $D_i$.
Now across block $m$ we are sampling $P(h_i(\xi))$ to generate $n_i$,
and then evaluating ${(\hat{h_i})}_{i,n_i}(x_i)$.  The associated
variances of values (one for each of the $x_i$),
Var$((\hat{h_i})_{i,\nu_i}(x_i))$, govern the accuracy of the estimate of
the free energy gradient.  For well-chosen $D_i$ these variances may
be far smaller than when $h_i = H$. In particular, if the number of
coordinates coupled to $\xi_i$ through ${H}$ does not grow as the
system does, often such difference Hamiltonian variances will not grow
much with system size, whereas the variances
Var$(\hat{H}_{i,\nu_i}(x_i))$ will grow greatly. Furthermore, very
often such a difference Hamiltonian is far easier to evaluate than is
$H$, due to cancellation in subtracting $D_i$.

More precisely, for practical reasons we want $i$'s update algorithm
to be robust against misperception of $q_{_{(i)}}$.  So for quadratic
loss, assuming no $\tau_{x_i} = 0$, consider the $h_i$ minimizing
$\int d{q_{_{(i)}}}$ $P(q_{_{(i)}}) [\int dn^u_i P(n^u_i \mid
n^{\xi_i}_i, q_{_{(i)}}, h_i)$ $\{{\vec{F}}^i_{H}(q(\xi)) -
{\hat{F}}^i_{n_i}(q_i)\}^2$], where $P(q_{_{(i)}})$ reflects any prior
information we might have concerning $q_{_{(i)}}$ (e.g., that it is
likely that the associated ${\vec{F}}^i_{H}(q)$ is close to that
estimated for the previous block of $\tau$ steps). Here
${\hat{F}}^i_{n_i}(q_i)$ is our estimator for ${\vec{F}}^i_{\beta
H}(q)$, and $n^u_i$ is the Hamiltonian values contained in $n_i$, the
associated $\xi_i$ values, $n_i^{\xi_i}$, being independent of $h_i$
and $q_{_{(i)}}$ and therefore fixed.

The inner integral is a sum, of the (square of the) bias
${\hat{F}}^i(q_i) - {\vec{F}}^i_{H}(q)$ with the variance, $\int
dn^u_i P(n^u_i \mid n^{\xi_i}_i, q_{_{(i)}}, h_i) \;
\{{\hat{F}}^i_{n_i}(q_i) - {\hat{F}}^i(q_i)\}^2$, where
${\hat{F}}^i(q_i) \equiv \int dn^u_i P(n^u_i \mid n^{\xi_i}_i,
q_{_{(i)}}, h_i) {\hat{F}}^i_{n_i}(q_i)$.  By only considering
difference utilities we guarantee that the bias equals 0. Now expand
our variance of vectors as a sum of variances of scalars, one for each
value $x_i$.  Since $n_i$ is IID generated that sum is $\frac{|\xi_i|
- 1}{|\xi_i|} \sum_{x_i} {\mbox{Var}}((\hat{h_i})_{i,\nu_i}(x_i)) /
\tau_{x_i}$.  The difference Hamiltonian minimizing this is ${H}(s) -
\sum_{x'_i} \frac{\tau_{x'_i}^{-1}}{\sum_{x''_i} \tau_{x''_i}^{-1}}
{H}(x'_i, {x}_{_{(i)}})$~\cite{wolp03a}.  (If all the terms in this
sum cannot be stored because $|\xi_i|$ is too large, then
$n_i^{\xi_i}$ must be averaged over as well, which can be approximated
by replacing the $\tau_{x_i}$ with $q(x_i)\tau$.)  Being independent
of $q_{_{(i)}}$, this Hamiltonian minimizes our $q_{_{(i)}}$ integral,
regardless of $P(q_{_{(i)}})$. For the same reason it is optimal if if
the integral is replaced by a worst-case bound over $q_{_{(i)}}$.

An extensive series of experiments have been conducted with this
optimal Hamiltonian under the approximation of uniform $q(x_i)$, and
under the approximation that $\tau_{x_i} = 0$ for one and only one
$x_i$ value.  These experiments compared this Hamiltonian with the
team game, for many different $H$, under a variant of the parallel
Brouwer update rule~\cite{wowh00, wotu02c} that crudely corrected for
non-stationarity. In that work these algorithms were semi-formally
justified as ways to minimize $H$, with no appreciation for free
energy, self-consistency, or the like. Indeed, due to the nonlinearity
of the Brouwer update rule's dependence on ${[h_i]}_{i,q^g}$, bad
distortions may arise with the update used in that previous work,
$q(x_i) \rightarrow e^{-\beta E(h_i \mid n_i, \xi_i = x_i)} /
\sum_{x'_i} e^{-\beta E(h_i \mid n_i, \xi_i = x_i)}$.  Also, there are
no known assurances that small Var$(h_i \mid n_i, \xi_i = x_i)$ result
in more accurate updating with parallel Brouwer updating, even when
that updating is done correctly.  Despite these shortcomings, in the
experiments in~\cite{wowh00, wotu02c} the approximations to the
optimal Hamiltonian minimized $H$ up to orders of magnitude faster
than team games, with the improvement growing with problem size.

\section{AVOIDING LOCAL MINIMA}

Say we have an $m=1$ team game, are ultimately interested in annealing
$\beta$, and are currently at a local minimum $q^{H}
\in \cal{Q}$ of the shared free energy. Then to break out of that
minimum we can simply raise $\beta$ and restart the updating, since we
want to raise $\beta$ anyway, and in general doing so will change the
${F}_{H}$ so that the Lagrange gaps become nonzero.

A way to break free without changing the free energies is to switch to a
coordinate system $\Xi^2$ for $\xi$, and thereby change $\cal{P}$.  As
an example, $\Xi^2$ can join two components of $\xi$ into an aggregate
coordinate. Since we can now have statistical dependencies between
those two components, the $\xi^2$ space product distributions map to a
superset of ${\cal{P}}$. In general the local minima of that superset
do not coincide with local minima of ${\cal{P}}$.

Less trivially, say $\xi^2 = \xi$, and $\xi(.)$ is the identity map
for all but a few components, indicated as indices $1 \rightarrow
n$. Have $\xi(.)$ be a bijection, so that for any fixed $x^2_{n+1
\rightarrow M} = x_{n+1 \rightarrow M}$, the effect of the coordinate
transformation is merely to ``shuffle'' the associated mapping taking
coordinates $1\rightarrow n$ to $\zeta$. Say we have a $m=1$ team
game, and set $q^{\xi^2}_{n+1 \rightarrow M} = q^{\xi}_{n+1
\rightarrow M}$. This means we can estimate the expectations of $\beta
H$ conditioned on possible $x^2_{1\rightarrow n}$ from the Monte Carlo
samples conditioned on $\xi(x^2_{1\rightarrow n})$. So for any
$\xi(.)$ we can estimate $E(H)$ as $\int dx^2_{1\rightarrow n}
p^{\xi^2}(x^2_{1\rightarrow n}) E(H \mid \xi(x^2_{1, \ldots,
n}))$. Now entropy is the sum of the entropy of coordinates $n+1
\rightarrow M$ plus that of coordinates $1 \rightarrow
n$. Accordingly, for any choice of $\xi(.)$ and
$q^{\xi^2}_{1\rightarrow n}$, we can approximate ${L}^{\xi}_{H}$
as (our associated estimate of) $E(H)$ minus the entropy of
$p^{\xi^2}_{1\rightarrow n}$, minus a constant unaffected by choice of
$\xi(.)$.

So for finite and small enough $|\xi_{1\rightarrow n}|$, we can use
our estimates $E(H \mid \xi(x^2_{1 \rightarrow n}))$ to search
for the ``shuffling'' $\xi(.)$ and distribution
$q^{\xi^2}_{1\rightarrow n}$ that minimizes ${L}^{\xi}_{H}$
(penalizing by the bias$^2$ plus variance expression if we intend to
do more Monte Carlo). The search can involve a series of free energy
descents over $\xi^2_{1\rightarrow n}$ for each possible $\xi(.)$, or
use cruder heuristics, like having $q^{\xi^2}_{1\rightarrow n} =
q^{\xi}_{1\rightarrow n}$, and only varying $\xi(.)$. Not only should
this coordinate transformation lower the free energy, it should also
result in a new surface through ${\cal{P}}^\dagger$ that is no longer
at a local minimum. More generally, for arbitrary $\xi^2$ we can bound
$F_{\xi^2} \ge F_{\xi^1} \ge F_{\xi^2} - {\mbox{max}}_{x^1}$
[ln$(\sum_{x^2} \delta_{\xi^1(x^2), x^1})]$ (and similarly for
uncountable $\xi^2$). So if after switching to $\xi^2$ we can then
reduce $F_{\xi^2}$ to less than the value $F_{\xi^1}$ had when we made
the switch, then we know we have also reduced $F_{\xi^1}$ to below
that pre-switch value. An upper bound on the how large that drop in
$F_{\xi^2}$ needs to be is ${\mbox{max}}_{x^1}$ [ln$(\sum_{x^2}
\delta_{\xi^1(x^2), x^1})]$.

$ $

\noindent
I would like to thank Chiu Fan Lee, Bill Macready, and Stefan
Bieniawski for helpful discussion.


\begin{thebibliography}{10}

\bibitem{auha92}
R.J. Aumann and S.~Hart.
\newblock {\em Handbook of Game Theory with Economic Applications}.
\newblock North-Holland Press, 1992.

\bibitem{chjo02}
D.~Challet and N.~F. Johnson.
\newblock Optimal combinations of imperfect objects.
\newblock {\em Phys. Rev. Let.}, 89:028701, 2002.

\bibitem{kono03_short}
G.~Korniss et~al.
\newblock {\em Science}, 299:677, 2003.

\bibitem{jayn57}
E.~T. Jaynes.
\newblock Information theory and statistical mechanics.
\newblock {\em Physical Review}, 106:620, 1957.

\bibitem{nish01}
H.~Nishimori.
\newblock {\em Statistical physics of spin glasses and information processing}.
\newblock Oxford University Press, 2001.

\bibitem{pean87}
C.~Peterson and J.~R. Anderson.
\newblock A mean-field theory learning algorithm for neural networks.
\newblock {\em Complex Systems}, 1:995, 1987.

\bibitem{pisl02}
E.~W. Piotrowski and J.~Sladkowski.
\newblock An invitation to quantum game theory.
\newblock /quant-ph/0211191, 2002.

\bibitem{wotu02c}
D.~Wolpert and K.~Tumer.
\newblock Beyond mechanism design.
\newblock In H.~Gao et~al., editor, {\em International Congress of
  Mathematicians 2002 Proceedings}. Qingdao Publishing, 2002.

\bibitem{wolp03a}
D.~H. Wolpert.
\newblock Theory of collective intelligence.
\newblock In K.~Tumer and D.~H. Wolpert, editors, {\em Collectives and the
  Design of Complex Systems}, New York, 2003. Springer.

\bibitem{wotu03a}
D.~H. Wolpert, K.~Tumer, and E.~Bandari.
\newblock Improving search by using intelligent coordinates.
\newblock 2003.
\newblock submitted.

\bibitem{wowh00}
D.~H. Wolpert, K.~Wheeler, and K.~Tumer.
\newblock Collective intelligence for control of distributed dynamical systems.
\newblock {\em Europhysics Letters}, 49(6), March 2000.

\end{thebibliography}
\end{document}